\documentstyle[12pt,psfig]{article}
\begin{document}
\begin{titlepage}
\begin{center}
{\large \bf {Hyperon-nucleon interactions in the
$p p \rightarrow K^+ \Lambda p$ reaction}}
\end{center}
\centerline{\bf N. G. Kelkar and B. K. Jain}
\begin{center}
{\it Nuclear Physics Division, 
Bhabha Atomic Research Centre, Mumbai 400 085, India.}
\end{center}

\begin{abstract}
We present calculations of the invariant mass spectra of the
$\Lambda$p system for the exclusive $p p \rightarrow K^+ 
\Lambda p$ reaction with the aim of studying the final state
interaction between the $\Lambda$-hyperon and the proton.
The reaction is described within a meson exchange framework
and the final state $\Lambda p$ interaction is incorporated
through an off-shell t-matrix for the $\Lambda p \rightarrow 
\Lambda p$ scattering, constructed using the available hyperon-nucleon (YN)
potentials. The cross sections are found to be sensitive to
the type of YN potential used especially at the 
$\Lambda$ and $\Sigma$ production thresholds.  
Hence, data on this exclusive reaction, which can be used to
constrain the YN potentials are desirable.  
\end{abstract}
\end{titlepage}

\newpage

\section{Introduction}
A good knowledge of the hyperon-nucleon (YN) interaction is crucial
for the understanding of a variety of phenomena ranging from
low energy hypernuclear physics to the strangeness production in
high energy heavy ion collisions. However, due to the non-availability
of sufficient YN scattering data, the YN interaction is still
quite unknown. 
Apart from a recent measurement of the differential cross sections
for $\Sigma ^+ p$ elastic scattering \cite{ahn}, 
only an almost three decades old total and differential 
cross section data on YN scattering 
from bubble chamber experiments
\cite{bubbl} exist. 
Since the experiments with short-lived
secondary beams of strange particles are difficult to perform,
reactions like the $K^- d \rightarrow  \pi^- Y N$
and $\pi^+ d \rightarrow K^+ Y N$, 
involving the production of a YN pair in the final state, 
have received much attention \cite{kdata}. 
Additionally, in the recent past, cross sections for 
the $p p \rightarrow K^+ Y N$
 reaction have been measured \cite{sieb,frasca} at the
Saturne National Laboratories
for proton beam energies of 2.3 and 2.7 GeV and at COSY for proton
energies very near to the threshold \cite{bale,cosy}.
Compared to the kaon and pion induced reactions, 
the $p p \rightarrow K^+ Y N$ reaction has the advantage that
(i) the proton beams are easily available and  
(ii) the data are free from uncertainties due to the deuteron target
as one can use a proton for the target nucleus.

Over the past few years, several theoretical 
models using different approaches 
(see ref.\cite{sibir} for a detailed review) like the
resonance model \cite{tsushima} and meson exchange model  
of Laget \cite{lage}
have tried to explain the total cross section data 
\cite{bale,landol,total} as well as the
missing mass spectra of ref.\cite{frasca}
on the $p p \rightarrow K^+ Y N$ reaction. The calculation
of the missing mass spectra by Deloff \cite{delo} 
using two different models of YN forces included only
the `direct kaon emission' (DKE) diagram and achieved 
qualitative agreement with data.  
Laget's calculation
involved a ($\pi$ + $K$) exchange mechanism in addition to DKE. 
With the cut-off mass 
$\Lambda_K$ in the off-shell form factor for the exchanged
kaon and the coupling constants $g_{N\Lambda K}$ 
and $g_{N\Sigma K}$ as free
parameters, this model achieved good quantitative
agreement with data.

However, the above analyses either neglect the final-state interaction 
between the $\Lambda p$ pair
\cite{sibir,tsushima} or use phenomenological prescriptions 
for the off-shell YN amplitudes. 
These approaches do not address the sensitivity
of the cross sections to the specific description 
of the hyperon-nucleon potential $V_{\Lambda p}$.
Due to the large momentum transfer involved in
the $p p \rightarrow K^+ \Lambda p$ reaction, we believe that the
cross sections for this reaction should be sensitive to the details
of the $\Lambda p$ potential.
In a recent study of the weak strangeness producing
reaction $p n \rightarrow p \Lambda$ \cite{parry} it was indeed
found that apart from the weak transition potential the cross sections
are very sensitive to the type of YN potential used. 
However, the cross sections for the weak reactions are very 
small ( $\sim 10^{-12}$ mb). Hence the 
$p p \rightarrow K^+ \Lambda p$ reaction which is experimentally
feasible is a better choice for the study of the $\Lambda p$ 
interaction.

Currently there exist meson exchange models of $V_{YN}$ 
which are similar in construction
to the models of the well-known nucleon-nucleon (NN) interaction.
The prominent ones among these are the Nijmegen and J\"ulich
potentials \cite{nijnnln,nij2,nij3,juln}. 
The free model parameters of the YN interaction in these 
 models are fixed using the cross section
data \cite{bubbl} on YN scattering. 
In addition to the meson exchange models, attempts to study
the YN interaction using quark models have also been made
\cite{qmodel1,qmodel2,qmodel3}. The $V_{YN}$ in these models
are constructed either using the resonating group method (RGM)
or the quark cluster model approach. In the 
RGM \cite{qmodel2}, the total quark
hamiltonian used consists of a confining piece, one gluon 
exchange potential and a chiral field induced quark-quark potential. 
The calculated YN phase shifts in this approach are found
similar to those due to Nijmegen model F. The cluster model
approach focuses mainly on the YN interactions at short distances
and explores their characteristics from the spin and flavour
symmetry structure, using one-gluon-exchange and the confining 
potential for the quark hamiltonian. In our work we use the 
boson-exchange models.

The aim of the present work is to study the sensitivity of the
missing mass spectra in the 
$p p \rightarrow K^+ \Lambda p$ reaction to  
$V_{YN}$, using as complete a description of the reaction as possible. 
We describe the $\Lambda p$ wave function in the
final state by using the solution of the 
Lippmann-Schwinger equation. The t-matrix for the 
$\Lambda p \rightarrow \Lambda p$ scattering in this equation is
constructed using the Nijmegen soft-core \cite{nijnnln} 
and J\"ulich \cite{juln}
potentials.  
This t-matrix includes the intermediate
$\Sigma$ state and hence the $\Lambda p$ wave function generated
by us includes the effect of the  
$\Sigma N$ channel on the $\Lambda p$ scattering.

As for the reaction mechanism, guided by the 
earlier investigations of the 
$p p \rightarrow K^+ \Lambda p$ reaction using $\pi$ and $K$ exchange
models \cite{sibir,lage}, we assume that it 
proceeds by the exchange of a pion
or a kaon between the two protons in the entrance channel and neglect
any contributions from heavier mesons like the $\rho$ and $K^*$. 
In the case of $\pi$ exchange, one of the interacting 
protons can be excited by the exchanged
pion to any of the relevant $N^*$ resonances, which then decays to a
$K^+$ and $\Lambda$. In the case of kaon exchange,  
the $K^+$ and $\Lambda$ are produced directly.
In addition to these meson exchange diagrams, we also include 
 the DKE diagram, where the proton in 
the initial state dissociates
into $K^+$ and an off-shell $\Lambda$, which by its interaction 
 with the second proton is brought on-shell. 
However its contribution is expected to be small
because the intermediate $\Lambda$ is {\it far} off-shell.

In our work we 
have included all the production diagrams shown in Fig. 1.
Since it is not possible to fix the relative sign between
the amplitudes corresponding to the pion and kaon exchange
diagrams we choose to 
retain an additive sign between these amplitudes. 
The diagrams arising due to antisymmetrization of the
two protons in the initial state have also been included.
The amplitudes for $\pi ^o p \rightarrow
 K^+ \Lambda$ and $K^+ p \rightarrow K^+ p$ at the
kaon production vertices are constructed from
the existing experimental data. Since these amplitudes fit
the experimental data, they implicitly include the excitation
of the resonances in $\pi p$ scattering. The off-shell 
extrapolation is incorporated through a form factor.

In section 2 we describe the formalism used to evaluate 
the transition amplitude for the 
$p p \rightarrow K^+ \Lambda p$ reaction. In section 3 we give the
results. 
Experimental data on the missing mass spectra of the
exclusive $p p \rightarrow K^+ \Lambda p$ reaction do
not exist. Hence we compare our results with the inclusive data
of ref.\cite{sieb} on the $p p \rightarrow K^+ Y N$ reaction.
Upto the $\Sigma$ threshold these data correspond only to the
missing mass spectra for the $p p \rightarrow K^+ \Lambda p$ 
reaction. However, above the $\Sigma$ threshold, the
contributions from the $p p \rightarrow K^+ \Sigma N$ channels 
take over.
The magnitude and shape of our calculated $\Lambda p$ mass spectra 
for the $p p \rightarrow K^+ \Lambda p$ reaction are found to be
sensitive to the type of $\Lambda p$ potential used. 
When compared with the inclusive data in the region below the
$\Sigma$ threshold, they are in reasonable agreement with the data.
However, the extent of agreement depends on the type of $\Lambda p$
potential used. 
At the opening of the $\Sigma$ threshold, results with  
the J\"ulich potentials show a 
pronounced peak which
appears only as a small kink in the case of the Nijmegen potential.
For a more stringent constraint on the
YN potentials, exclusive data on the various channels in 
the $p p \rightarrow K^+ Y N$ reaction are required.

In addition to the study of the $\Lambda p$ interaction we
have also studied some other aspects of the
$p p \rightarrow K^+ \Lambda p$ reaction.
Instead of using the detailed description for the $\Lambda p$
final state interaction as done in the present paper, it is 
always tempting to approximate it by an on-shell t-matrix
multiplied by an off-shell extrapolating form factor, as
done in ref. \cite{lage}. We have investigated the extent of
accuracy of such a procedure by comparing the off-shell t-matrix
computed by us with the corresponding on-shell t-matrix multiplied
by an off-shell form factor.  
A detailed discussion of these results and those on the 
relative contributions of the various diagrams of Fig. 1
are given in section 3.
Finally in section 4 we give a summary of this work.

Though this work has the limited plan of studying the 
$p p \rightarrow K^+ \Lambda p$ reaction, 
we plan to extend our calculations to calculate the cross sections
for the $\Sigma$ producing channels too in future.

\section{Formalism}
The differential cross section for the reaction $p + p \rightarrow
K^+ + p + \Lambda$ is given as \cite{kaja},
\begin{equation}\label{1}
d\sigma\, =\, {1 \over F} \,\, \, \int {d\vec{p_K}\over 2E_K}\,\,
{d\vec{p_3}\over 2E_3}\, \,{d\vec{p_{\Lambda}}\over 2E_{\Lambda}}\,\,
\,\delta(P_i-P_f)\,\,\,\biggl<\,\,\biggl|\,T_{fi}\,\biggr|^2\,\,\biggr>
\end{equation}
where F is the {\it flux factor} and the angular brackets around
$|T_{fi}|^2$ denote the average and sum over the spins of the
particles in the initial and final states respectively.
The momenta and energies of the two protons in the initial state
and the $K^+$, $\Lambda$ and proton in the final state are denoted
respectively as, $\vec{p_1}$, $\vec{p_2}$, 
$\vec{p_K}$, $\vec{p_{\Lambda}}$, $\vec{p_3}$ and $E_1$, $E_2$, $E_K$, 
 $E_{\Lambda}$, $E_3$.
Expressing the phase space integral in the above equation in terms of 
variables in the $\Lambda p$ centre of mass system (CMS), the double
differential cross section is given as,
\begin{equation}
{d^2\sigma \over d\Omega_K dW}\, = \,[PS]\,\,\times \,\, \int\, 
d\Omega^{(\Lambda p)}_{\Lambda} \,\, 
\biggl<\,\,\biggl|\,T_{fi}\,\biggl|^2\,\,\biggr>
\end{equation}
where $\Omega_K$ is the laboratory solid angle of the emitted $K^+$.
W is the invariant mass of the $\Lambda p$ pair.
$\Omega_{\Lambda}^{(\Lambda p)}$ is   
the solid angle of the $\Lambda$ in the $\Lambda p$ CMS and 
PS is the phase space factor. All the kinematics is done 
relativistically.

$T_{fi}$, within a meson exchange framework (Fig. 1) consists of,
\begin{equation}\label{3}
T_{fi} =  T_K + T_{\pi} + T_{DKE}
\end{equation}
where $T_K$ and $T_{\pi}$ correspond to the amplitudes of the kaon and 
pion exchange diagrams of Figs. 1a and 1b and $T_{DKE}$ is the
amplitude for direct $K^+$ emission as shown in Fig. 1c.
The DKE by the proton leads to the formation
of an off-shell $\Lambda$ which is brought on-shell only 
after interacting
with the second proton.

In what follows, we shall
describe the various ingredients required for evaluating $T_{fi}$.
Detailed expressions using partial
wave expansions for the t-matrices being lengthy, 
are given in the appendix.
\vskip0.5cm

\centerline{ {\bf A. Final state interaction}} 
\vskip0.5cm

Each of the T-matrices in eq.(\ref{3}), which includes the interaction
between the $\Lambda$ and proton, is given by,
\begin{equation}\label{4}
T_x \,=\, \biggl <\,\Psi_{\Lambda p}^-\,\,\vec{p_K}\,\,m_{\Lambda}\,m_3
\,\biggl|\,V^x_{p p \rightarrow K^+ \Lambda p}\,\biggl|\,\,
\vec{p_1}\,\,\vec{p_2}\,m_1\,m_2\,\biggr > \,-\, exchange \,\,term
\end{equation}
where $x$ is $\pi$, K or DKE.
$m_1$, $m_2$, $m_{\Lambda}$ and $m_3$ are the spin projections
of the two entrance channel protons and the final state $\Lambda$ and proton
respectively. The exchange term arises due to the antisymmetrization
of the two entrance channel protons and is written by interchanging
$\vec{p_1}$, $m_1$ with $\vec{p_2}$, $m_2$ in the first
term of the above equation. 
The final state $\Lambda p$  
 wave function $\Psi^{-*}_{\Lambda p}$ consists of a plane wave and
a scattered wave and can be written as,
\begin{equation}\label{5}
\Psi^{-*}_{\Lambda p} \,=\, \biggl <\,\vec{p_{\Lambda}}\,\vec{p_3}\,
\biggl |\,\,\,\, +\,\,\,\, \Psi^*_{scat}
\end{equation}
where $\Psi^*_{scat}$ is given in terms of the t-matrix for $\Lambda p$ 
scattering as,
\begin{equation}\label{6}
\Psi^*_{scat}\, = \, <\,\vec{p_{\Lambda}}\,\vec{p_3}\,\biggl|\,
\,t_{\Lambda p \rightarrow \Lambda p} \,\,\,G_o\,.
\end{equation}
Here $G_o$ is the plane wave propagator for the $\Lambda p$ system in
the intermediate state. Since the $\Lambda p$ momentum in this state
is not fixed, the matrix $t_{\Lambda p \rightarrow \Lambda p}$ in the
above expression is necessarily off-shell. The phase shift approximation
to $\Psi_{scat}$ is obtained by taking this t-matrix on-shell. In
r-space this implies approximating the scattered wave function by its 
asymptotic form.

To obtain $\Psi_{scat}$, we consider the $\Lambda N$ as well as the
$\Sigma N$ channels together.  
$\Psi_{scat}$ is thus obtained by
solving, 
\begin{equation}\label{7}
t_{\Lambda p \rightarrow \Lambda p}\, = \, V_{\Lambda p 
\rightarrow \Lambda p}\,\, + \,\,
\biggl < \,\,\Lambda \,p\,\,\biggl|\,\,V\,\,G\,\,t\,\,\biggl |\,
\Lambda \, p\,\biggr> 
\end{equation}
where V, G and t are $(2 \times 2)$ matrices. G is the free
$\Lambda p$ or $\Sigma N$ propagator between two scatterings.
The diagonal matrix 
elements describe respectively the $\Lambda p \rightarrow \Lambda p$
and $\Sigma N \rightarrow \Sigma N$ channels while the off-diagonal
elements give the $\Lambda p \rightarrow \Sigma N$ and $\Sigma N
\rightarrow \Lambda p$ transitions. Thus the constructed 
$t_{\Lambda p}$ includes the effect of the $\Sigma N$ channel in
$\Lambda p$ scattering. 
We construct $t_{\Lambda p}$ using the Nijmegen soft-core 
\cite{nijnnln} and
J\"{u}lich \cite{juln} YN potentials. We present results with both, 
the energy dependent versions (A, B) and energy independent versions
(\~{A}, \~{B}) of the J\"ulich potential. 
The Nijmegen and J\"ulich models A and \~{A} 
are single meson exchange
models whereas the J\"ulich models B and \~{B} 
involve higher order
processes involving the intermediate $\Delta$ and $Y^*$. 
Both the Nijmegen and J\"ulich models 
provide a similar description of the
YN total cross section data. However, there are significant differences
in the inputs of these models. They differ in the choice of meson 
parameters, the contribution of the various mesons to the YN 
interaction, their spin structure 
and in the treatment of non-localities. 
A detailed comparison of the various YN models
can be found in refs. \cite{juln,reub}.  
\vskip0.3cm
\centerline{{\bf B. Born amplitude $V^x_{p p 
\rightarrow K^+ \Lambda p}$}}
\vskip0.5cm
The Born amplitude $V^x_{p p \rightarrow K^+ \Lambda p}$  
of eq. (\ref{4}) is constructed
using the kaon and pion exchange diagrams of Figs. 1a and 1b and the
direct kaon emission (DKE) diagram in Fig. 1c. The DKE requires the
Lagrangian for the $pK^+\Lambda$ vertex, which for the pseudoscalar
coupling is given by,
\begin{equation}\label{8}
{\cal L}_{pK^+\Lambda}\,=\, i \,g_{K\Lambda N}\,\, \bar{\Psi}_{\Lambda}
\,\,\gamma_5\,\,\,\Psi_p\,\,
\vec{\tau} \cdot \,\vec{\phi_K}
\end{equation}
where $\vec{\tau}$ is an isospin operator at the 
$p K^+ \Lambda$ vertex. 
The expectation value of this Lagrangian
over p and $\Lambda$ wave functions is
\begin{eqnarray}\label{12}
\biggl <\,\vec{q_{\Lambda}}, \,\Lambda \,\biggl |
{\cal L}_{pK^+\Lambda}\,\biggl |\,p,\,\vec{p_1}\,\biggr> &=&  
i \,g_{K\Lambda N}\,
 \Biggl [{(E_1+M_p)\,(E_{\Lambda} + M_{\Lambda})
\over 2\,M_p\,\,2\,M_{\Lambda}} \Biggr ]^{1 \over 2}\, \\ \nonumber
&\times& \biggl <\, \,\Lambda \,\biggl |\, 
\vec{\sigma}
\cdot \, \Biggl [\,{\vec{p_1} \over E_1 + M_p} \,-\, {\vec{q_{\Lambda}}
\over E_{\Lambda} + M_{\Lambda}} \biggr ] \, 
\biggl |\,\,p\,\biggr> \, \, \vec{\tau} \cdot \vec{\phi_K}
\end{eqnarray}
In addition to ${\cal L}_{pK^+\Lambda}$, the kaon exchange diagram 
involves the elastic scattering $K^+ p \rightarrow K^+ p$ of an 
off-shell kaon on a proton. A useful approximation to this amplitude
which includes most of the dynamics at the scattering vertex, is
to use a scattering amplitude determined phenomenologically from
$K^+ p$ elastic scattering and corrected for the off-shell 
extrapolation by a form factor, $F_K(q_K^2)$ where $q_K$ is the 
momentum carried by the off-shell kaon. For our calculations
we take the scattering amplitude from ref. \cite{tkaon} and use
the monopole form, $F_K(q_K^2)=(\Lambda^2_K - M_K^2) /(\Lambda^2_K
- q_K^2)$ for the form factor.

For the pion exchange diagram we need the Lagrangian 
${\cal L}_{\pi N N}$ and the amplitude for the $\pi^o p 
\rightarrow K^+ \Lambda$ process. Like in the case of 
${\cal L}_{pK^+\Lambda}$ we use the pseudoscalar coupling for
${\cal L}_{\pi N N}$. The on-shell 
matrix elements for $\pi^o p \rightarrow K^+ \Lambda$ 
 are related to those for the $\pi^-p \rightarrow K^o
\Lambda$ reaction (for which the experimental data is available)
due to isospin symmetry. We use the partial wave amplitudes given
in ref.\cite{tpion} which were constructed from all available
experimental data on $\pi^-p \rightarrow K^o \Lambda$.
The off-shell correction to this amplitude is again made using
a monopole form factor, $F_{\pi}(q_{\pi}^2)$ where $q_{\pi}$ is
the momentum carried by the off-shell pion.

In order to describe consistently the Born transition 
amplitude $V^x_{p p \rightarrow K^+ \Lambda p}$ 
and the $\Lambda p$ potentials
which we use for the final state interaction,
we take the values of the coupling constants and 
cut-off masses in the transition amplitude identical to those used 
in the construction of the $\Lambda p$ potentials \cite{juln}.
The values of $g_{K \Lambda N}$ and 
$g_{\pi N N}$ are
taken to be -14.1 and 13.3 respectively. The cut-off masses 
$\Lambda_K$ and $\Lambda _{\pi}$ in the monopole form factors
$F_K(q_K^2)$ and $F_{\pi}(q_{\pi}^2)$ are taken to be 1.2 and
1.3 GeV respectively.

All our calculations are done using the partial wave expansion
at every stage (see appendix for detailed expressions). Thus 
from the calculation point of view, our results
are exact and carry all the details of the final state interaction
and the amplitude $V^x_{p p \rightarrow K^+ \Lambda p}$.

In this work we do not include the contribution to the $p p 
\rightarrow K^+ \Lambda p$ reaction from the two step process
which initially involves the excitation of a proton in the
entrance channel to $\Sigma$ and then its conversion to $\Lambda$ 
through $\Sigma N \rightarrow \Lambda N$. 
The cross section for
this process was found to be very small in the calculation 
of Ref. \cite{lage}. It was also found to be negligible in
Ref. \cite{haid} where the authors studied the 
weak production of the $\Lambda$ using a coupled channel approach
for the $\Lambda p$ final state interaction. However, in a 
recent calculation \cite{gasp} of the total cross sections
for the $p p \rightarrow K^+ \Lambda p$ reaction near threshold,
a rather large contribution of this process was reported. 
In view of the results of Refs. \cite{lage,haid} and considering
the smallness of 
the coupling constant involved at the $N\Sigma K$ vertex as compared
to that at the $N\Lambda K$ vertex, we do not expect a significant
contribution from the above mentioned process.

\section{Results and Discussion}
The main objective of our calculations is to study 
the effect of the final state
interaction between the proton and $\Lambda$ in the $p p \rightarrow
K^+ \Lambda p$ reaction and make an inter comparison of  
the calculated cross sections using the different available models
of the hyperon-nucleon (YN) interaction. The
differences in the various models of the YN interaction are 
expected to show up in the $\Lambda$p invariant mass spectra 
because they involve $\Lambda$p at different relative momenta
starting from zero at threshold. These spectra should therefore 
be sensitive to the differences in the contributions
of the various partial waves in different potentials 
for the $\Lambda p$ scattering 
and to the opening of the $\Sigma$
production channel where the $^3S_1-\,^3D_1$ coupled channels are
important.

In Fig. 2 we show the differential cross section 
$d^2\sigma/d\Omega _KdW$ as a function of the invariant mass W 
of the $\Lambda p$ pair. $\Omega_K$ is the solid angle of the
emitted kaon in the laboratory system. The proton beam 
energy is 2.3 GeV and the kaon
angle is fixed at 10$^o$. 
The thick solid line in Fig. 2a is our
full calculation for the $p p \rightarrow K^+ \Lambda p$ reaction  
with the Nijmegen potential for the $\Lambda p$
interaction. By `full' we mean that the calculation 
includes all diagrams of Fig. 1, namely 
$\pi$ plus K exchange and $K^+$ direct emission. The
dash-dotted and dashed lines are respectively the full 
calculations with the energy
independent versions \~{A} and \~{B}  of the J\"ulich
potentials. 
The thin solid line is the calculation using plane waves for the
$\Lambda$ and proton. In Fig. 2b we show the calculations using the
energy dependent versions A and B of the J\"ulich potentials. 
We observe that all the curves with the 
$\Lambda p$ interaction included show an enhancement of the
cross sections over the
plane wave results, along the entire mass spectrum. The extent
of the enhancement and the detailed structure in the spectrum,
however, depend upon the choice of $V_{YN}$.

We now discuss the structures appearing in the $\Lambda p$
invariant mass spectra. 
As can be seen in Fig. 2a, the J\"ulich potential \~{A} produces
a prominent cusp at the $\Sigma$ threshold which
is less prominent in the case of the Nijmegen potential.
A rounded peak in the cross section at the $\Sigma$ threshold is
produced by the J\"ulich potential \~{B}. This peak is shifted
in position as compared to the cusps. 
Similarly, at the 
$\Lambda$ threshold we see sharp peaks or bumps in Figs. 2a and 2b,
depending on the type of potential used. 
The shapes of these structures at the $\Lambda$ and $\Sigma$
thresholds, as shown recently in Ref. \cite{miyag}, are 
related to the location
of the YN t-matrix poles in the various partial waves. 
The locations of these poles themselves, depend on the details of 
the YN potential used. All the above mentioned potentials
differ in their weightage of the different partial waves, 
though they have been constructed
by fitting to the same set of $\Lambda N$ elastic scattering data.

Next, we discuss the individual contributions
of the various partial waves to the cross sections.
Since the S waves are the most
dominant in $\Lambda p \rightarrow \Lambda p$ scattering, in Fig. 3
we show the cross sections obtained by omitting the interactions
in the S wave channels and the coupling between the $^3S_1$
and $^3D_1$ channel.
The calculations are done using all the diagrams of Fig. 1 and two
types of hyperon-nucleon potentials, namely the Nijmegen and the
J\"ulich \~{A}. The thin solid line is the plane wave calculation
in which the final state interaction of the $\Lambda$ and proton
is switched off completely. As can be seen in the figure, 
in the case of the Nijmegen $\Lambda p$ potential, removal of the
interaction in the $^3S_1-\,^3S_1$ channel brings down the
cross sections (see dashed line) almost to the plane wave results. 
Omitting the
$^3S_1-\,^3D_1$ and $^3D_1-\,^3S_1$ transitions in
$\Lambda p \rightarrow \Lambda p$ scattering (dash-dotted line) 
has a small effect and the removal of the $^1S_0-\,^1S_0$ channel
causes a negligible change in the cross sections and
cannot be seen in the figure.
The J\"ulich potential however has its strength distributed more evenly
over the various channels. The $^1S_0-\,^1S_0$ transition
specifically in this potential seems to be responsible for the peak 
at the $\Lambda$ threshold.

Let us now see how the calculated cross sections using the various
$\Lambda p$ potentials compare with the available data.
In Fig. 4 we compare our calculated cross sections with the inclusive
data of ref.\cite{sieb} on the $p p \rightarrow K^+ Y N$ reaction,
since there is no exclusive data available on the missing mass
spectra in the $p p \rightarrow K^+ \Lambda p$ reaction.
Below the threshold for the $\Sigma$
production (W = 2128 MeV) the data correspond entirely to that for the
$p p \rightarrow K^+ \Lambda p$ reaction. The steep rise in the
measured cross sections at the $\Sigma$ threshold is due to
the opening of the $\Sigma$ producing channels
$p p \rightarrow K^+ \Sigma^o p$ and 
$p p \rightarrow K^+ \Sigma^+ n$. 
We see that, while both the Nijmegen and J\"ulich potentials
give the general trend of the data below the $\Sigma$ threshold,
in details they compare differently. 
The cross sections calculated using J\"ulich \~{A} agree
somewhat with data at the $\Lambda$ threshold. 
The peak at the $\Sigma$ threshold
is more pronounced with the J\"ulich potentials than with the
Nijmegen potential.
To constrain the YN potentials further, we need exclusive data
on the $\Lambda p$ and $\Sigma N$ channels.

Next, we investigate the accuracy of the on-shell approximation
to the $\Lambda p$ final state interaction. 
For this we use the Nijmegen YN potential. 
In Fig. 5, the thick solid line is our calculation with the
off-shell $\Lambda p$ t-matrix as in Fig. 2. The thin solid line
is obtained by replacing the off-shell matrix elements
$<J\,M\,(L S)\,p_{\Lambda p}\, 
 |\,t^{\Lambda p \rightarrow \Lambda p}
\, |\,q\,(L^{\prime} S^{\prime})\,J\,M>$ in our calculations by 
the on-shell ones $<J\,M\,(L S)\,p_{\Lambda p}\, 
 |\,t^{\Lambda p \rightarrow \Lambda p}\, 
|\,p_{\Lambda p}\,(L S)\,J\,M>$. The scale for this
curve is indicated on the right hand side of the figure. 
The scale for all other curves is indicated on the left.
We see that the on-shell approximation results are much larger
in magnitude and different in shape compared to those due to the
correct t-matrix. This difference, however, is reduced to a great
extent by multiplying the on-shell t-matrix by an off-shell
form factor. The results obtained by 
repeating the same on-shell calculation with 
 the above t-matrix elements multiplied by a form
factor $F_{\Lambda p} = (p_{\Lambda p}^2 + \beta^2)/(q^2 + \beta^2)$ as
in ref.\cite{lage} with $\beta = 1.36 fm^{-1}$ 
are shown by the dashed curve of Fig. 2.  
The form factor causes a large reduction
of the cross sections and the pronounced peak at the $\Lambda$ threshold
flattens out. Still a considerable difference between this
prescription and the correct calculation persists. This difference
can now be reduced and the calculated ``on-shell'' results can be
brought nearer to the experimental measurement, by arbitrarily
adjusting the parameters in the reaction vertices for the 
$p p \rightarrow K^+ \Lambda p$ transition. The dash-dotted curve
in Fig. 2 which agrees with data is the result for 
$\Lambda _{\pi}$, $\Lambda _K$ and $g_{KN\Lambda}$ equal to
1.1 GeV, 0.925 GeV and -13.26 respectively. However, 
this way of constraining the transition
amplitude parameters is obviously misleading.

So far we have discussed the results using the reaction mechanism
involving all the diagrams of Fig. 1. Let us now see the
contributions of the individual diagrams with the pion and kaon
exchange and direct $K^+$ emission. In Fig. 6 we plot the 
cross sections with only pion exchange (dashed curves), only
kaon exchange (dash-dotted curves) and the full calculation with
$\pi$ plus $K$ exchange and $K^+$ direct emission (solid curves) 
for the Nijmegen and J\"ulich \~{A} $\Lambda p$ interactions. 
The contribution of the direct $K^+$ 
emission diagram is found to be negligibly small.  
The contribution of the kaon exchange diagram is much larger than that
of the pion exchange diagram but it does not account completely
for the cross sections. 
The dominance of the
kaon-exchange diagram found here is consistent with some recent
results reported by the DISTO Collaboration \cite{disto}.
They performed a measurement of the polarization transfer
coefficient $D_{NN}$ for the $\vec{p} p \rightarrow p K^+ 
\vec{\Lambda}$ reaction and found it to be very large and
negative which is consistent with a mechanism dominated by 
kaon-exchange. 
In the next section we summarize the findings of this work.

\section{Summary}
We have studied the hyperon-nucleon (YN) final state interactions (FSI)
in the $p p \rightarrow K^+ \Lambda p$ reaction at 2.3 GeV beam
energy. The invariant $\Lambda p$ mass spectra are calculated within
a meson exchange framework. The FSI between the $\Lambda$ and the proton 
are incorporated through a t-matrix for $\Lambda p \rightarrow \Lambda
p$ scattering. This t-matrix is constructed using the Nijmegen and J\"ulich 
YN potentials. Due to unavailability of data on the invariant mass
spectra in the exclusive $p p \rightarrow  K^+ \Lambda$ p reaction we
compare our results with data on the inclusive $p p \rightarrow 
K^+ Y N$
reaction. Upto the $\Sigma$ threshold the inclusive
data receives contribution only from the exclusive 
$p p \rightarrow K^+ \Lambda p$ channel.

The cross sections with FSI included are found to be enhanced compared to
the plane wave results for both the potentials. 
However, the magnitude of the cross sections and the structure in them
differ a lot for the two potentials. 
The general trend of the experimental data is produced by these 
calculations but in details they compare differently with data. 
Thus the exclusive data on the $p p \rightarrow K^+ \Lambda p$ reaction
can be very useful to differentiate amongst different $V_{YN}$.

Regarding the reaction mechanism involved, we find that the kaon exchange
diagram gives the dominant contribution to the cross sections. This is
consistent with some recent measurements of $D_{NN}$ for the 
$\vec{p} p \rightarrow p K^+ \vec{\Lambda}$ 
reaction made by the DISTO collaboration.

We have made a comparison of our results with the
off-shell t-matrix for $\Lambda p \rightarrow \Lambda p$ scattering with
those using the on-shell t-matrix. The on-shell results (even after
multiplying by the off-shell form factor) overestimate the results
obtained using the off-shell $\Lambda p$ t-matrix. Therefore, attempts
to represent the FSI between the $\Lambda$ and proton by phase shifts
(with an off shell form factor) may not represent the FSI accurately.  
This can lead to errors in conclusions about the parameters associated
with the $p p \rightarrow K^+ \Lambda p$ transition interaction, if the
same are adjusted to fit the data in this approach.

The present work can be improved by taking into account the actual
off-shell nature of the $K N \rightarrow K N$ and $\pi N \rightarrow
 K \Lambda$ T-matrices. 
The final state $K \Lambda$ interaction may
not be as small as that expected for the $K N$ pair. It is 
incorporated in some sense in the pion exchange terms of the
cross sections in this work since we use 
$T_{\pi N \rightarrow K \Lambda}$ constructed from experimental data.
However it needs to be included explicitly in the kaon-exchange
terms too. A calculation of the total
cross sections and polarization observables with the $K \Lambda$
interaction included is planned for the future. 

\vskip0.5cm
\centerline {{ \bf ACKNOWLEDGEMENTS}}
The authors wish to thank Angels Ramos for computational help
related to the $\Lambda p \rightarrow \Lambda p$ t-matrix and for
some useful discussions.

\newpage
\setcounter{equation}{0}
\centerline{\bf APPENDIX: TRANSITION AMPLITUDE FOR 
$p p \rightarrow K^+ \Lambda p$}
\renewcommand{\theequation}{A.\arabic{equation}}
Let us start by writing the transition amplitude for the direct terms
using the various diagrams shown in Fig. 1. The amplitudes for the exchange
terms arising due to the antisymmetrization of the two initial
protons are written by interchanging $\vec{p_1}$, $E_1$, $m_1$ with
$\vec{p_2}$, $E_2$, $m_2$ in the equations to follow. 
In a distorted wave Born approximation the transition amplitude
for the $p p \rightarrow K^+ \Lambda p$ reaction is given as,
\begin{equation}\label{14}
T_x = \sum_{S M_S} \,\biggl 
< {1 \over 2} \, {1 \over 2}\, m_3 \,m_{\Lambda}\, 
\biggl |\, S\, M_S \,\biggr > \,\,\biggl <\vec{p_K}\, 
\Psi^-_{\vec{p_{\Lambda p}} \,S\, M_S} 
\,\biggl | \, V^x_{pp\rightarrow K^+\Lambda p}\,  \biggl |\, 
\vec{p_1}\, \vec{p_2}\, m_1 \,m_2 \biggr > \nonumber
\end{equation}
where $x$ is either $\pi$, $K$ or DKE corresponding to pion and 
kaon exchange and direct $K^+$ emission as shown in Fig. 1 and
$\vec{p_{\Lambda p}}$ is the momentum of the $\Lambda$ and 
proton in the $\Lambda p$ centre of mass system. 
$\Psi^-_{\vec{p_{\Lambda p}} S M_S}$ is the distorted relative 
wave function
for the $\Lambda p$ pair in the final state. 
$< \vec{p_K} |$ is the relative plane
wave of $K^+$ with respect to the centre of mass of the 
$\Lambda p$ system. 
Now using the Lippmann-Schwinger
equation for $\Psi^-_{\vec{p_{\Lambda p}} S M_S}$ we get,
\begin{eqnarray}\label{15}
&&
T_x=\sum_{S M_S} \biggl < {1 \over 2} 
{1 \over 2} m_3 m_{\Lambda}
\biggl | S M_S \biggr > 
\biggl [ 
\biggl <\vec{p_K} \vec{p_{\Lambda p}} S M_S  
\biggl | V^x_{p p \rightarrow K^+ \Lambda p} \biggl |
\vec{p_1}\vec{p_2} m_1 m_2 \biggr > \\ \nonumber
&+&\sum_{S^{\prime} M_S^{\prime}}
\int d\vec{q} {\biggl<\vec{p_{\Lambda p}} S M_S \biggl |
t^
{\Lambda p \rightarrow \Lambda
p}(\omega)\biggl |\vec{q} S^{\prime}M_S^{\prime}\biggr >\over 
\omega - E(q) + i \eta }
\biggl < \vec{p_K} \vec{q} S^{\prime} M^{\prime}_S
 \biggl | V^x_{p p \rightarrow K^+ \Lambda p}
 \biggl |
\vec{p_1} \vec{p_2} m_1 m_2 \biggr > \biggr ] 
\end{eqnarray}
where $\omega$ is the energy corresponding to the asymptotic 
momentum $\vec{p_{\Lambda p}}$.
The two terms in the square bracket correspond to the two 
diagrams in each of the Figs. 1a and 1b.

In the case of DKE (Fig. 1c), the amplitude $
V^x_{p p \rightarrow K^+ \Lambda p}$ reduces to the Lagrangian 
${\cal L}_{p K^+ \Lambda}$ and the $\Lambda$ is produced off-shell with
a momentum $\vec {q_{\Lambda}} = \vec{p_1} - \vec{p_K}$. Since the
observed $\Lambda$ is on-shell, the first term in 
the square bracket of eq. (A.2) does not
contribute. It is brought on-shell by its interaction with the other
proton only, as given in the second term. The amplitude for DKE is
thus given as,
\begin{eqnarray}\label{25}
T_{DKE}\, = 
 i \, g_{K\Lambda N}
\,\, \sqrt{{E_1 + M_p \over 2 M_p}}\,\,
 \sqrt{{E(q_{\Lambda}) + M_{\Lambda} \over 2 M_{\Lambda}}}\,
\,\sum_{S M_S}\,\biggl <\,{1\over 2}\, {1\over 2}
\,m_3 \, m_{\Lambda}\, \biggl |\, S\, M_S\, \biggr > \\ \nonumber
\times \sum_{S^{\prime} M_S^{\prime}\,m_{\Lambda}^{\prime}}{\biggl 
<\,\vec{p_{\Lambda p}}\, S \, M_S
\,  \biggl |\,t^{\Lambda p \rightarrow \Lambda p}(\omega) \,
 \biggl | \,\vec{q_{\Lambda p}}\, S^{\prime}\, M_S^{\prime} \,
\biggr > \over \omega - E(q_{\Lambda p})} \, \,\, 
\biggl <\,{1 \over 2}\,{1 \over 2}\,m_2\,m_{\Lambda}^{\prime}\,
 \biggl |\,S^{\prime} \,M_S^{\prime}\,\biggr > \\ \nonumber
\,\times\,\,\,\,\biggl <\,m_{\Lambda}^{\prime}\, \biggl |
\,\vec{\sigma} \cdot \biggl ( {\vec{p_1} \over E_1 + M_p} \,-\,
 {\vec{q_{\Lambda}} \over E(q_{\Lambda}) + M_{\Lambda}} 
\biggr )\, \biggl | \, m_1 
\biggr > 
\end{eqnarray}   
where $\vec{q_{\Lambda p}}$ is the momentum of the proton and 
off-shell $\Lambda$ in the $\Lambda p$ centre of mass system.


The matrix elements of the plane wave amplitude  
$V^K_{p p \rightarrow K^+ \Lambda p}$ for kaon exchange are 
given as,
\newpage
\begin{eqnarray}\label{17}
\biggl < \vec{p_K} \vec{p_3} \vec{p_{\Lambda}} m_3 
\,m_{\Lambda}
 \biggl | V^K_{p p \rightarrow K^+ \Lambda p} \biggl |
\vec{p_1} \vec{p_2} m_1 m_2 \biggr > = 
{i  g_{K\Lambda N}
\over q_K^2 - M_K^2}\, \sqrt{{E_2 + M_p \over 2 M_p}}\,
 \sqrt{{E_{\Lambda} + M_{\Lambda} \over 2 M_{\Lambda}}}\\ \nonumber
\biggl 
<\,m_{\Lambda}\, \biggl |
\,\vec{\sigma} \cdot \biggl ( {\vec{p_2} \over E_2 + M_p} \,-\,
 {\vec{p_{\Lambda}} \over E_{\Lambda} + M_{\Lambda}} \biggr )
\, \biggl | \, m_2 
\biggr > 
\biggl [\,F_K(q_K^2\,)
\biggr ]^2 \\ \nonumber
\times \,\,\biggl<\,{1 \over 2}\, m_3 \, \vec{p_K}\, \vec{p_3}\,  \biggl |
\, T_{K^+p\rightarrow K^+p} \,  \biggl | \, {1 \over 2}\, m_1 \,
\vec{q_K} \, \vec{p_1}\, \biggr > \, \, \,\,\,\,
\end{eqnarray}
where $M_p$, $M_{\Lambda}$ and $M_K$ are the masses of the proton, 
 $\Lambda$ and kaon respectively,
$q_K$ is the four momentum carried by the exchanged kaon and the coupling
constant $g_{K\Lambda N}$=-14.1. 
The matrix elements
$<\,\, |T_{K^+p\rightarrow K^+p} |\,\,>$ are expanded in partial waves
and expressed in terms of phase shifts \cite{tkaon}. 
The monopole form factor 
$F_K(q_K^2)=(\Lambda^2_K - M_K^2) /(\Lambda^2_K
- q_K^2)$ (with $\Lambda _K$=1.2 GeV) takes
care of the off-shell nature of the exchanged kaon.

In a similar way, the 
 plane wave matrix elements for the pion exchange transition amplitude
(Fig. 1b) are given as,
\begin{eqnarray}\label{18}
\biggl < \vec{p_K} \,\vec{p_3}\, \vec{p_{\Lambda}} \,m_3 
\,m_{\Lambda}
 \biggl | V^{\pi}_{p p \rightarrow K^+ \Lambda p} \biggl |
\vec{p_1}\, \vec{p_2}\, m_1 \,m_2 \biggr > = {i \, g_{\pi NN}
\over q_{\pi}^2 - M_{\pi}^2}\,\, \sqrt{{E_2 + M_p \over 2 M_p}}\,\,
 \sqrt{{E_3 + M_p \over 2 M_p}}\, \\ \nonumber
\times\,\,\biggl <\,m_3\, \biggl |
\,\vec{\sigma} \cdot \biggl ( {\vec{p_2} \over E_2 + M_p} \,-\,
 {\vec{p_3} \over E_3 + M_p}\biggr ) \, \biggl | \, m_2 
\biggr > \, 
\biggl [\,F_{\pi}(q_{\pi}^2\,)\, \biggr ]^2 \\ \nonumber
\times \,\biggl<\,{1 \over 2}\, m_{\Lambda} \, \vec{p_K}\, 
\vec{p_{\Lambda}}\,  \biggl |
\, T_{\pi^op\rightarrow K^+\Lambda} \,  \biggl | \, {1 \over 2}\, 
m_1 \, \vec{q_{\pi}} \, \vec{p_1}\, \biggr > 
\end{eqnarray}
where $q_{\pi}$ is the four momentum
carried by the exchanged pion, $M_{\pi}$ its mass, the coupling constant 
$g_{\pi NN}$=13.3 and $F_{\pi}(q_{\pi}^2$) (with $\Lambda _{\pi}$=1.3 GeV
) is the monopole form factor which takes care 
of the off-shell nature of the
exchanged pion.  
The matrix elements $< |T_{\pi^o p \rightarrow K^+ \Lambda}| >$
are also expanded in partial waves and are written using the amplitudes
given in ref.\cite{tpion}.
We use the same form factors at both the upper and lower 
vertices in Figs. 1a and 1b for the kaon and pion exchange diagrams.

\vskip0.4cm
{\centerline { \bf Coupled channel method for $t^{\Lambda p \rightarrow 
\Lambda p}$}}
\vskip0.4cm
Since the mass difference between the $\Lambda$ and $\Sigma$ hyperon
is not large, the $\Sigma N$ channel plays an important role
in $\Lambda p$ scattering and should be treated in an exact coupled
channel method.
The matrix elements of the t-matrix in a coupled channel formalism
can be written in matrix form as, 
\begin{eqnarray}\label{19}
 &&\biggl <\,\vec{p_{\Lambda p}} S M_S
 |{\bf t}(\omega) |
\vec{q}\,S^{\prime} M_S^{\prime}\biggr > = \biggl 
< \vec{p_{\Lambda p}}\,S M_S\,\biggl |{\bf V}(\omega)
\biggl |\vec{q}\, S^{\prime}\,M_S^{\prime}\,\biggr > \\ \nonumber
&&+\,\,\sum_{S^{\prime \prime} M_S^{\prime \prime}}\,\int
k^2\,dk\,\biggl < \vec{p_{\Lambda p}}\,S M_S \biggl |
{\bf V}(\omega)\biggl | \vec{k}\,S^{\prime \prime}
M_S^{\prime \prime}
\biggr >\,{\bf G}_o(k,\omega)\, 
 \biggl <\vec{k}\,S^{\prime \prime} M_S^{\prime \prime}
\biggl |{\bf t}(\omega)\,\biggl |\vec{q}\,S^{\prime} M_S^{\prime}
 \biggr >
\end{eqnarray}
with,
\[ {\bf t}(\omega)\,\,=\,\,
\left(
\begin{array}{cc}
t^{\Lambda p \rightarrow \Lambda p} & t^{\Lambda p \rightarrow \Sigma N}\\
t^{\Sigma N \rightarrow \Lambda p} & t^{\Sigma N  \rightarrow \Sigma N}
\end{array}
\right )  
 \,\,\, \,\,\,\,\,,\,\,{\bf V}(\omega)\,\,=\,\,
\left(
\begin{array}{cc}
V^{\Lambda p \rightarrow \Lambda p} & V^{\Lambda p \rightarrow \Sigma N}\\
V^{\Sigma N \rightarrow \Lambda p} & V^{\Sigma N  \rightarrow \Sigma N}
\end{array}
\right ) \] \\
and,
\[ {\bf G}_o(k,\omega)\, \,=\,\, 
\left( 
\begin{array}{cc}
  \frac {1}{\omega - E_{\Lambda}(k) + i \epsilon } & 0 \\
0 &  \frac{1}{\omega - E_{\Sigma}(k) + i \epsilon }
\end{array}
\right ) \] \\
The matrix elements of ${\bf t}(\omega)$ 
can be expressed in terms of partial wave expansions
as,
\begin{eqnarray}\label{16}
&&\biggl <\, \vec{p_{\Lambda p}}\, S\, M_S\,  \biggl | \, 
{\bf t}(\omega)\, \biggl | \,
\vec{q} \, S^{\prime}\, M_S^{\prime}\, \biggr > = 
\sum_{J\,M}\, \sum_{L\,L^{\prime}\,M_L\,M_L^{\prime}}\,Y_{L M_L}
(\hat{p_{\Lambda p}})
\,Y_{L^{\prime} M_L^{\prime}}^*(\hat{q}) \\ \nonumber
\,&&\biggl < \,L\, M_L
\,S M_S\,  \biggl | J M \biggr >
\biggl <L^{\prime}  M_L^{\prime} S^{\prime} M_S^{\prime} 
\biggl |  J M \biggr > 
\,\biggl <J M (L S) p_{\Lambda p}
 \biggl |{\bf t}(\omega) 
\biggl |q\,(L^{\prime}\,S^{\prime})\,J M
\biggr >
\end{eqnarray}

We evaluate the matrix elements of $t^{\Lambda p \rightarrow 
\Lambda p} (\omega)$
in eq.(\ref{15}) numerically, using eqs.(\ref{19}) and (\ref{16}) and
the available YN potentials ${\bf V}(\omega)$. 
The on-shell amplitudes evaluated using our 
t-matrix code are in good agreement with the values
published in the original works of the Nijmegen and J\"ulich groups.

\newpage
\begin{figure}
\centerline{\vbox{
\psfig{file=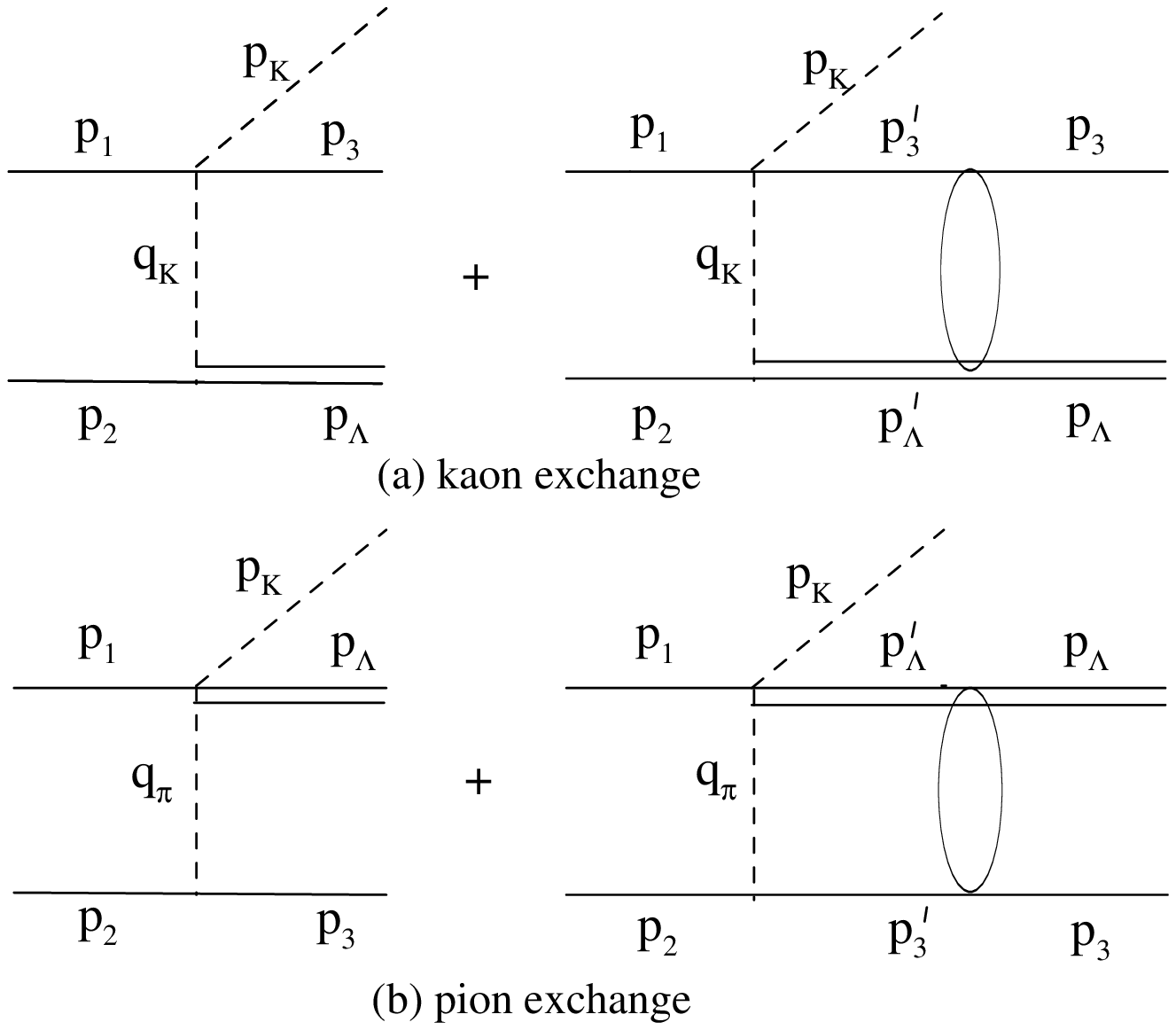,height=11cm,width=11cm}}}
\end{figure}
\begin{figure}
\centerline{\vbox{
\psfig{file=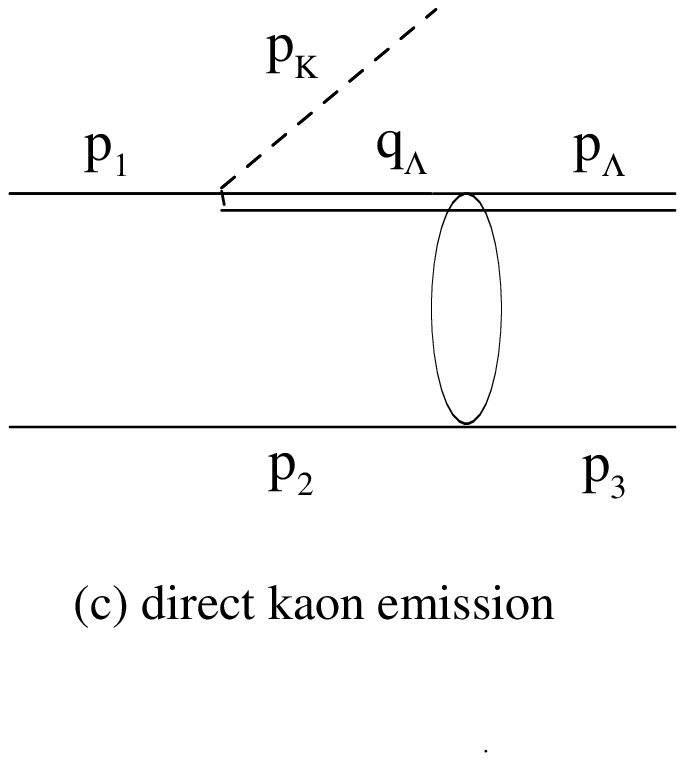,height=7cm,width=6cm}}}
\caption{
Diagrams for the $p p \rightarrow K^+ \Lambda p$ reaction
 (a) kaon exchange (b) pion exchange (c) direct kaon emission.
The ellipses indicate the final state interactions of the $\Lambda$
and proton.}
\end{figure}

\newpage
\begin{figure}
\centerline{\vbox{
\psfig{file=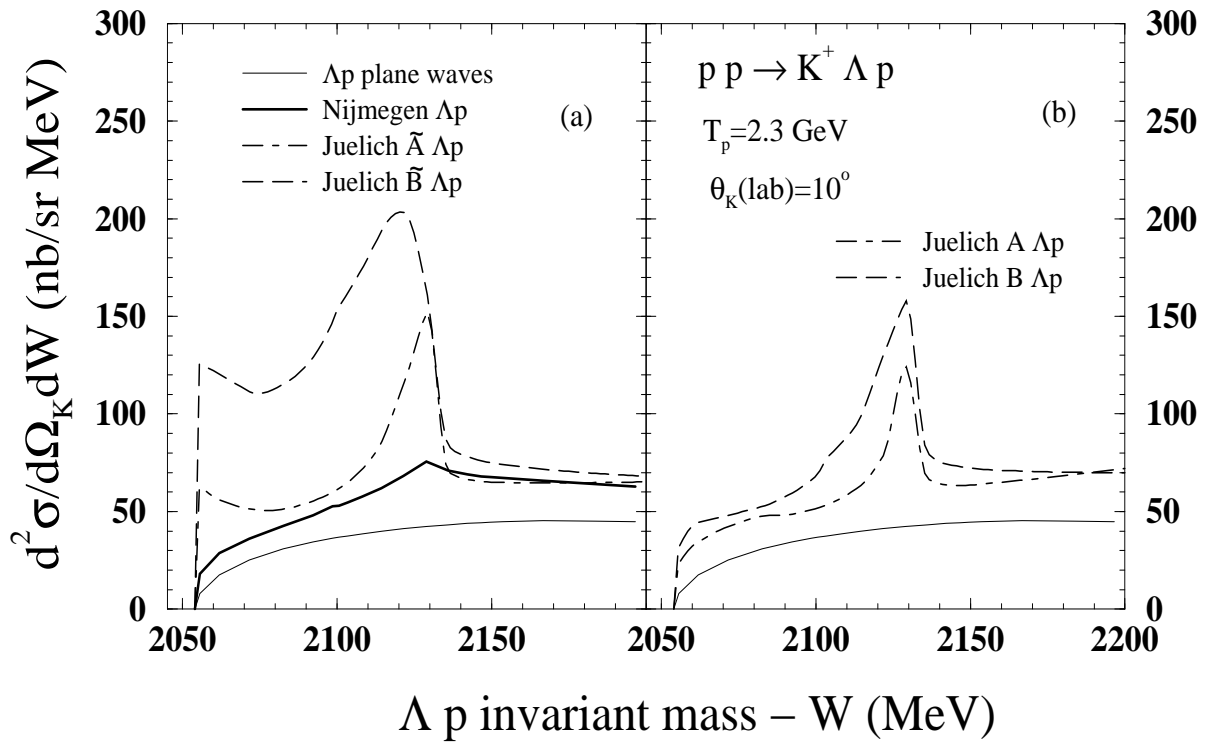,height=10cm,width=16cm}}}
\caption{
$\Lambda p$ invariant mass spectra at 2.3 GeV beam energy 
for the exclusive $p p \rightarrow
K^+ \Lambda p$ reaction calculated using different potentials for
the $\Lambda p$ final state interaction. (a) thick solid line uses
the Nijmegen soft-core potential, dash-dotted and dashed
curves are calculations using versions \~{A} and \~{B} 
respectively of the energy
independent J\"ulich potential. (b) Calculations using energy
dependent versions A (dash-dotted curve) and B (dashed curve) of the
J\"ulich potential. The thin solid lines in (a) and (b) are the
results using plane waves for the $\Lambda$ and proton in the 
final state.}
\end{figure}

\newpage
\begin{figure}
\centerline{\vbox{
\psfig{file=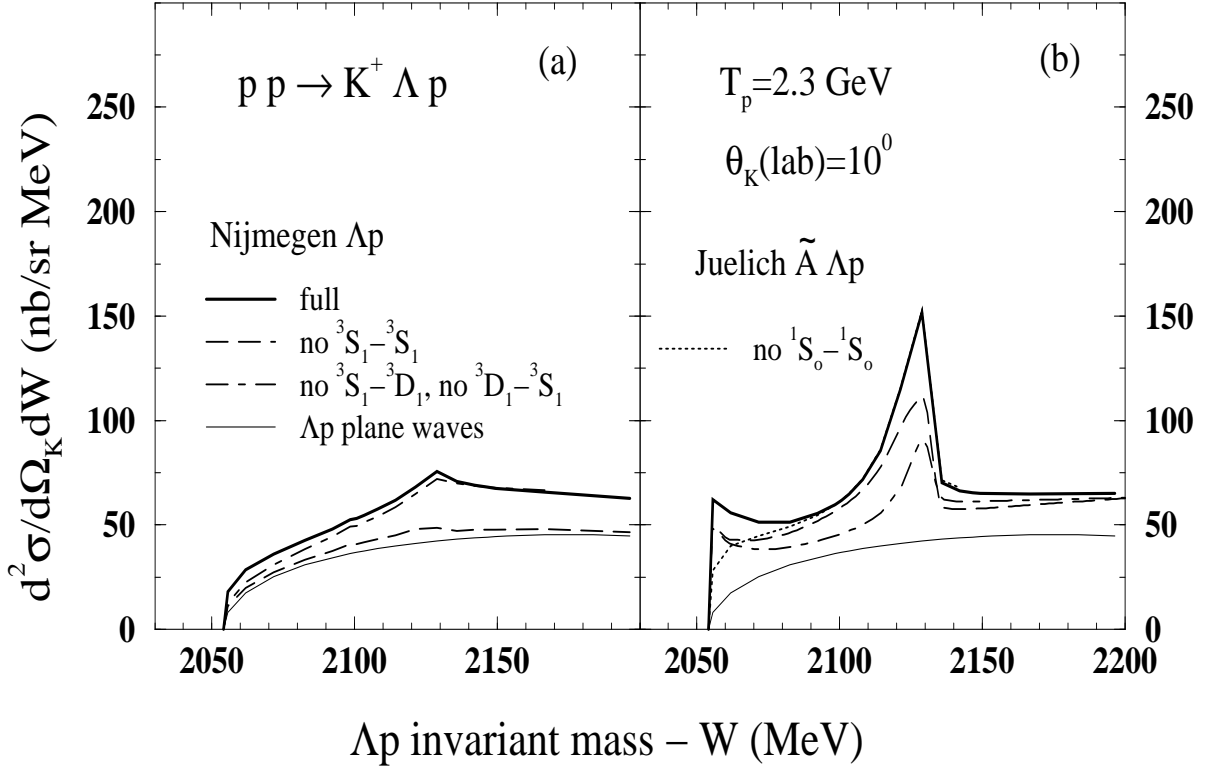,height=10cm,width=16cm}}}
\caption{
Contribution of the various channels 
of the $\Lambda p$ elastic scattering to the $\Lambda p$
invariant mass spectra for the $p p \rightarrow K^+ \Lambda p$ reaction. 
Thin solid lines are the cross sections obtained using plane waves
for the final state $\Lambda$ and proton. Dashed lines are the results
obtained after removing the $^3S_1-\,^3S_1$ channel from 
$<JM(LS)p_{\Lambda p} |t^{\Lambda p
\rightarrow \Lambda p}|q(L^{\prime}S^{\prime})JM>$.  
Dash-dotted lines are
the results obtained by dropping the $^3S_1-\,^3D_1$ and $^3D_1-\,^3S_1$ 
channels and dotted lines are those after removing the $^1S_0-\,^1S_0$
transition. Results are shown using two types of YN potentials
(a) Nijmegen and (b) J\"ulich \~{A}.}
\end{figure}

\newpage
\begin{figure}
\centerline{\vbox{
\psfig{file=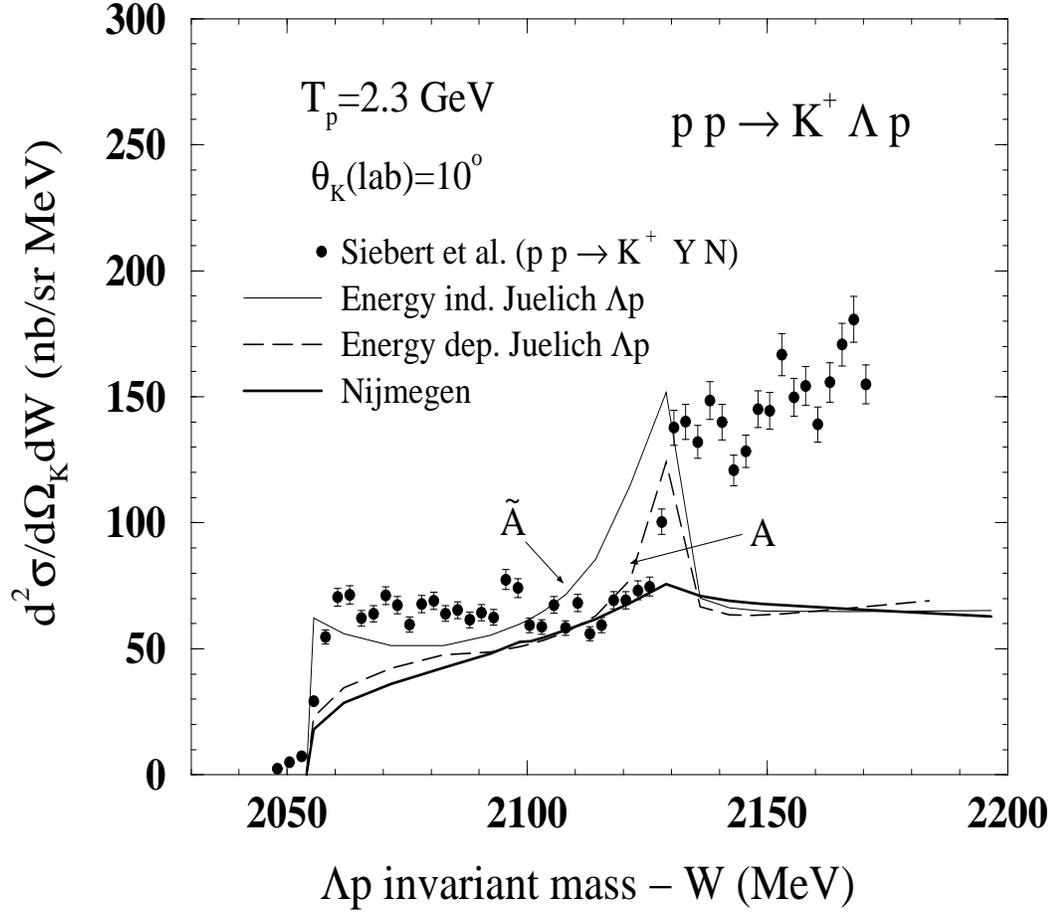,height=12cm,width=14cm}}}
\caption{
Comparison of calculated invariant $\Lambda p$ mass spectra
for the $p p \rightarrow K^+ \Lambda p$ reaction with the available
inclusive data on the $p p \rightarrow K^+ Y N$ reaction. The data
are from ref. \protect \cite{sieb}. The thick solid line is the 
calculation using the Nijmegen soft-core $\Lambda p$ potential. The 
thin solid line and dashed line are respectively the results using energy
independent and dependent versions of the J\"ulich potential A.}
\end{figure}

\newpage
\begin{figure}
\centerline{\vbox{
\psfig{file=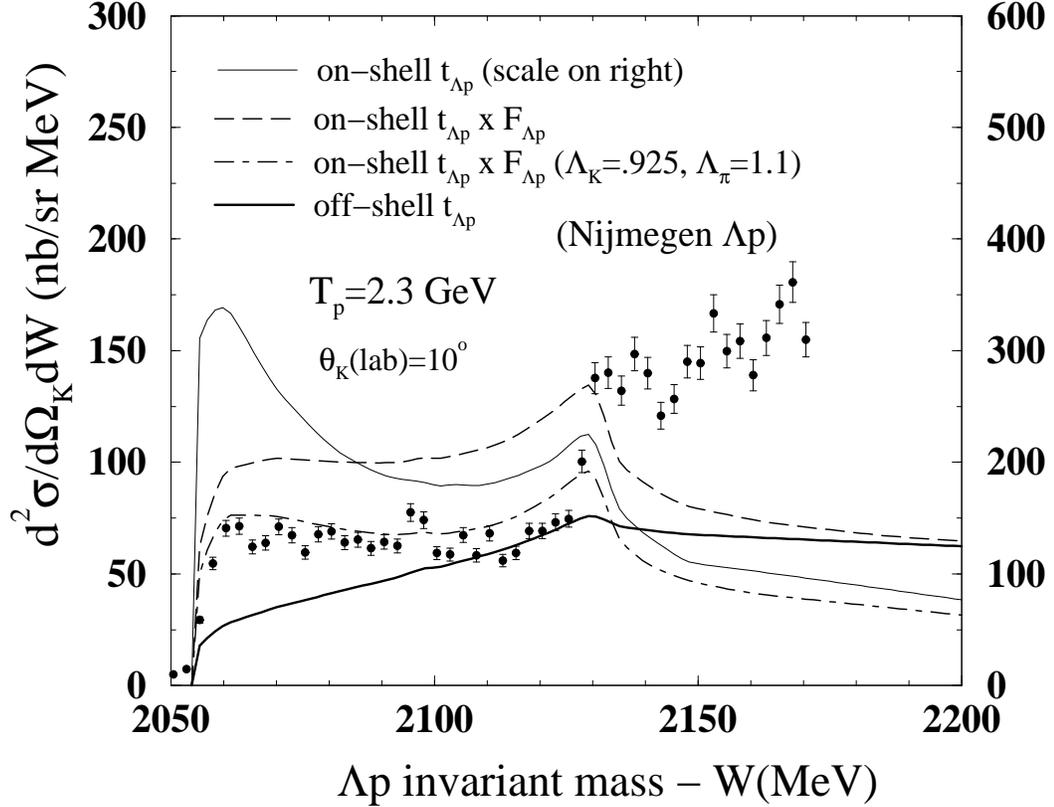,height=11cm,width=14cm}}}
\caption{
Effects of on-shell approximations of the final state $\Lambda p$
interaction on the $\Lambda p$ mass spectra for the $p p \rightarrow 
K^+ \Lambda p$ reaction. Calculations are done using the Nijmegen
$\Lambda p$ potential.  
The thin solid line is the calculation (with scale
given on the right)
using on-shell matrix elements of the t-matrix 
for $\Lambda p \rightarrow \Lambda p$. Dashed
curve is obtained by multiplying the on-shell matrix elements
of $t^{\Lambda p \rightarrow
\Lambda p}$   
by a form factor $F_{\Lambda p}$ (scale on left). 
Dash-dotted curve is obtained
by reducing the values of $\Lambda _K$ and $\Lambda _{\pi}$ in a calculation
similar to that of the dashed curve. The thick solid line uses the 
off-shell matrix elements of $t^{\Lambda p \rightarrow \Lambda p}$ 
(as used throughout this work). 
 The solid curves and the dashed curve show results
using $\Lambda_K =1.2$ and $\Lambda_{\pi}=1.3$ GeV. The data is the same
as in Fig. 4.}
\end{figure}

\newpage
\begin{figure}
\centerline{\vbox{
\psfig{file=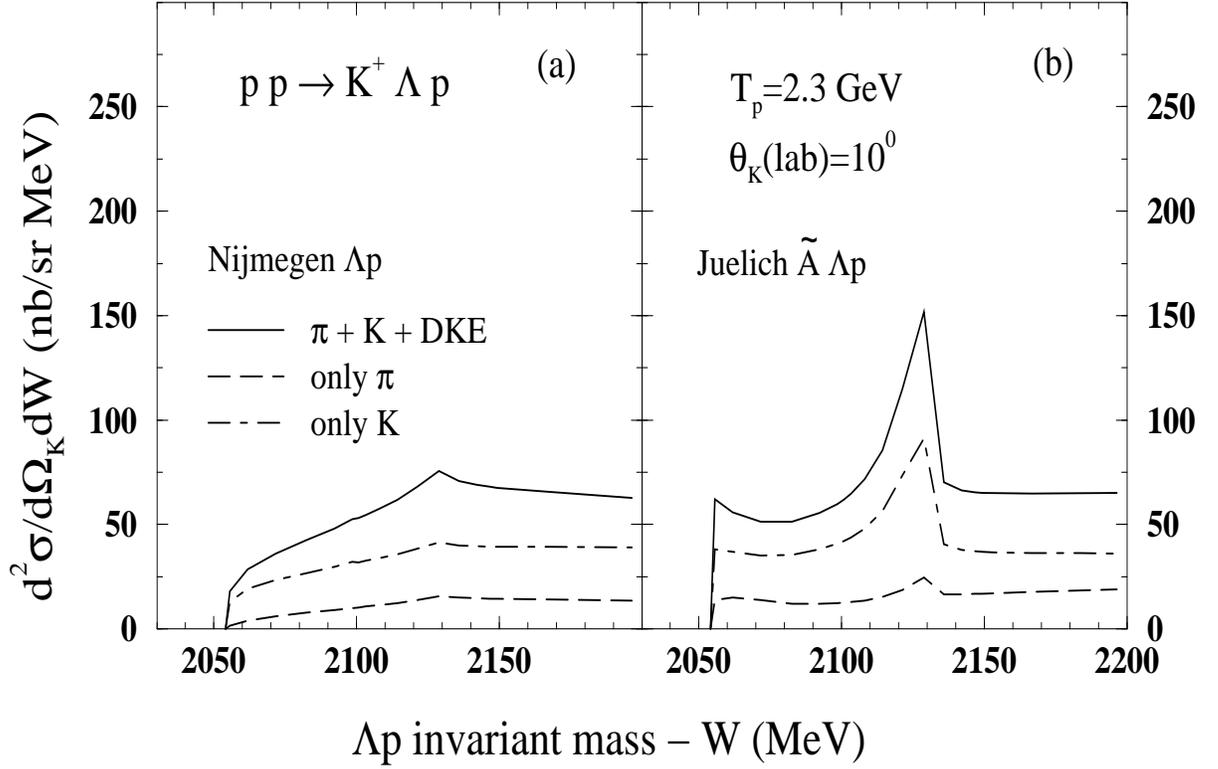,height=10cm,width=16cm}}}
\caption{
Contributions of the various meson exchange diagrams to
the $p p \rightarrow K^+ \Lambda p$ reaction using (a) Nijmegen
and (b) J\"ulich \~{A} potentials for $\Lambda p$ final state
interaction. The solid curves are the calculations including all
diagrams of Fig. 1. Dashed and dash-dotted curves are the cross sections
using only pion and only kaon exchange mechanisms respectively.}
\end{figure}

\end{document}